\documentclass[conference]{IEEEtran}
\IEEEoverridecommandlockouts
\usepackage{cite}
\usepackage{amsmath,amssymb,amsfonts}
\usepackage{algorithmic}
\usepackage{graphicx}
\usepackage{textcomp}
\usepackage{xcolor}
\usepackage{multirow}
\usepackage{setspace}

\def\BibTeX{{\rm B\kern-.05em{\sc i\kern-.025em b}\kern-.08em
    T\kern-.1667em\lower.7ex\hbox{E}\kern-.125emX}}

\makeatletter
\newcommand{\linebreakand}{%
    \end{@IEEEauthorhalign}
    \hfill\mbox{}\par
    \mbox{}\hfill\begin{@IEEEauthorhalign}
}
\makeatletter

\DeclareRobustCommand*{\IEEEauthorrefmark}[1]{%
    \raisebox{0pt}[0pt][0pt]{\textsuperscript{\footnotesize\ensuremath{#1}}}}

\begin{document}

\title{Leveraging Pre-trained Models for FF-to-FFPE Histopathological Image Translation
\thanks{\IEEEauthorrefmark{*}Corresponding author.}}

\author{
    \IEEEauthorblockN{Qilai Zhang\IEEEauthorrefmark{1}, Jiawen Li\IEEEauthorrefmark{1}, Peiran Liao\IEEEauthorrefmark{1}, Jiali Hu\IEEEauthorrefmark{2}, Tian Guan\IEEEauthorrefmark{1}, Anjia Han\IEEEauthorrefmark{3,*} and Yonghong He\IEEEauthorrefmark{1,4, *}}
    \IEEEauthorblockA{\IEEEauthorrefmark{1} Shenzhen International Graduate School, Tsinghua University, Shenzhen, China}
    \IEEEauthorblockA{\IEEEauthorrefmark{2} Medical Optical Technology R\&D Center, Research Institute of Tsinghua, Guangzhou, China}
    \IEEEauthorblockA{\IEEEauthorrefmark{3} Department of Pathology, The First Affiliated Hospital of Sun Yat-sen University, Guangzhou, China}
    \IEEEauthorblockA{\IEEEauthorrefmark{4} Jinfeng Laboratory, Chongqing, China}
    \IEEEauthorblockA{Email: \{zhang-ql22, lijiawen21, lpr22\}@mails.tsinghua.edu.cn, maggie\_0225@126.com, \\ guantian@sz.tsinghua.edu.cn, hananjia@mail.sysu.edu.cn, heyh@sz.tsinghua.edu.cn}
}

\maketitle

\begin{abstract}
The two primary types of Hematoxylin and Eosin (H\&E) slides in histopathology are Formalin-Fixed Paraffin-Embedded (FFPE) and Fresh Frozen (FF). FFPE slides offer high quality histopathological images but require a labor-intensive acquisition process. In contrast, FF slides can be prepared quickly, but the image quality is relatively poor. Our task is to translate FF images into FFPE style, thereby improving the image quality for diagnostic purposes. In this paper, we propose Diffusion-FFPE, a method for FF-to-FFPE histopathological image translation using a pre-trained diffusion model. Specifically, we utilize a one-step diffusion model as the generator, which we fine-tune using LoRA adapters within an adversarial learning framework. To enable the model to effectively capture both global structural patterns and local details, we introduce a multi-scale feature fusion module that leverages two VAE encoders to extract features at different image resolutions, performing feature fusion before inputting them into the UNet. Additionally, a pre-trained vision-language model for histopathology serves as the backbone for the discriminator, enhancing model performance. Our FF-to-FFPE translation experiments on the TCGA-NSCLC dataset demonstrate that the proposed approach outperforms existing methods. The code and models are released at https://github.com/QilaiZhang/Diffusion-FFPE.
\end{abstract}


\section{Introduction}
Histopathological Hematoxylin and Eosin (H\&E) slides are primarily prepared in two ways: Formalin-Fixed Paraffin-Embedded (FFPE) and Fresh Frozen (FF). FFPE, the standard in pathology, involves a lengthy preparation process of 24–48 hours \cite{rogers1987accuracy}, providing excellent glandular and cellular preservation but unsuitable for rapid intraoperative diagnosis. Conversely, FF slides are produced by freezing tissues in approximately 15 minutes, making them ideal for rapid surgical diagnosis and treatment planning \cite{jaafar2006intra}. However, FF preparation often leads to tissue fragility and ice crystal artifacts, which can impair diagnostic clarity \cite{trejo2015accuracy}. With advancements in deep learning, particularly generative networks, cross-domain style transfer now enables the transformation of FF slides to FFPE-like quality. This technique has significant potential to enhance the readability of digital pathology images, supporting faster and more accurate intraoperative diagnoses \cite{falahkheirkhah2022generative}.

The goal of FF-to-FFPE histopathological image translation is to transform FF images to the FFPE style while preserving original content. Due to the lack of pixel-matched FF and FFPE data pairs, unpaired image translation methods are necessary. Existing approaches \cite{ozyoruk2022deep, fan2022fast, li2023st} primarily use GANs to translate FF to FFPE images, emphasizing histopathological structure preservation and inference efficiency. However, they generally require training from scratch, demanding large datasets to achieve robust generalization \cite{ozyoruk2022deep}.


Generative models like Stable Diffusion have recently demonstrated strong capabilities in image generation \cite{rombach2022high}. These pre-trained models capture general image features effectively, enabling adaptability across domains. Fine-tuning them for pathology images leverages embedded prior knowledge to capture the distinct textures and structures of histopathology \cite{zhang2023customized}. The rise of pre-trained histopathology vision models has also significantly enhanced performance in downstream tasks \cite{lu2024visual}, and using these models as discriminator backbones further benefits GAN training \cite{kumari2022ensembling}.

Building on the above concept, we propose Diffusion-FFPE, a method for FF-to-FFPE histopathological image translation that leverages pre-trained models. This approach utilizes a pre-trained generative model as the generator and a pre-trained histopathology visual model as the discriminator, optimized through adversarial objectives to fully exploit embedded prior knowledge. Inspired by img2img-turbo \cite{parmar2024one}, we adopt a one-step diffusion model as the generator and fine-tune it using LoRA adapters \cite{hu2021lora}. To further enhance performance, we use CONCH \cite{lu2024visual} as the backbone of discriminator. Additionally, we introduce a multi-scale feature fusion module to capture the global structures (e.g., tissue contours) and fine details (e.g., nuclei) in histopathological images comprehensively.

\begin{figure*}[htbp]
    \centering
    \includegraphics[width=1\textwidth]{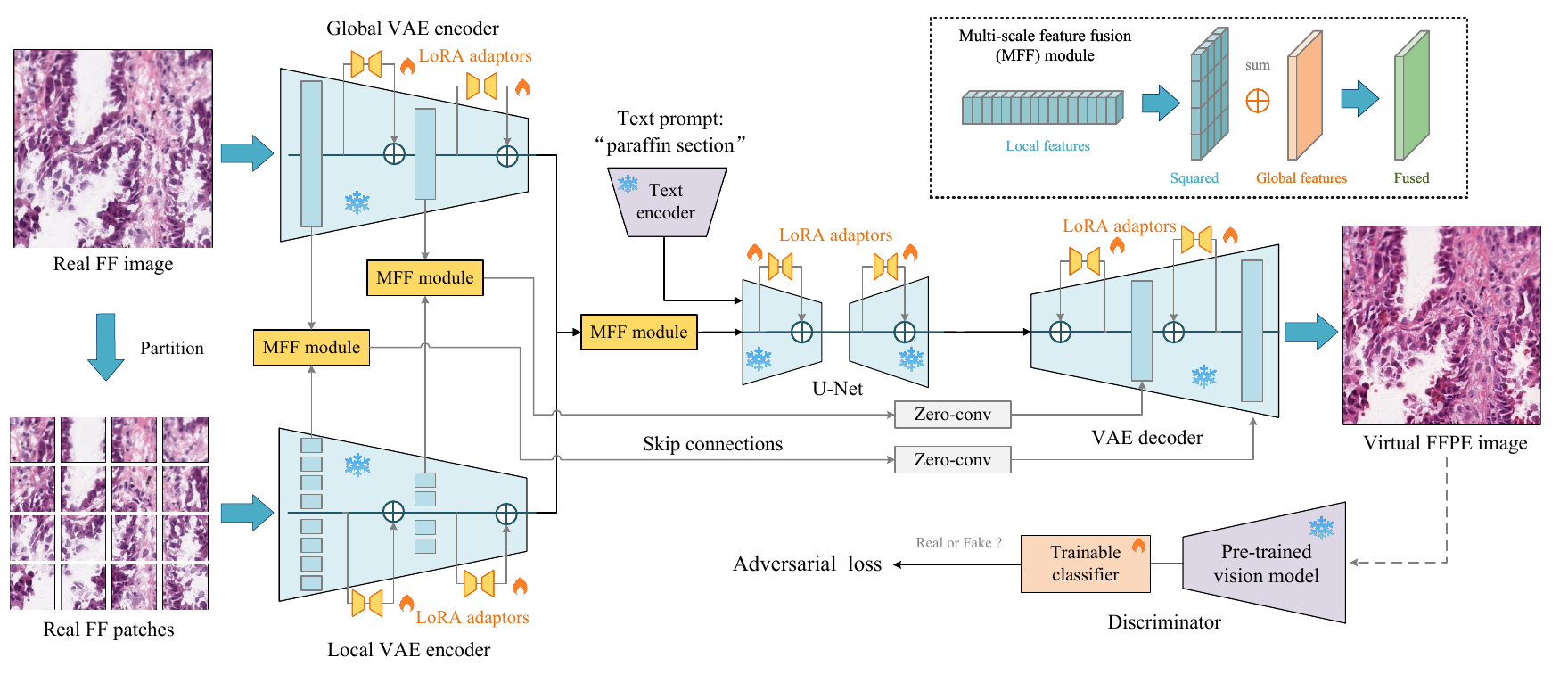}
    \caption{Overview of Diffusion-FFPE. During training, the generator's weights are fixed, with trainable LoRA adapters added to each of its components. The discriminator utilizes a pre-trained vision model as its backbone, followed by a trainable classifier. Intermediate features, fused by the MFF module from both the global and local VAE encoders, are forwarded to the VAE decoder through skip connections.}
    \label{overview}
\end{figure*}

In summary, the contributions of this paper are as follows:
\begin{itemize}
\item This paper proposes Diffusion-FFPE, a method that leverages pre-trained models for both the generator and discriminator for FF-to-FFPE histopathological translation.
\item We propose a multi-scale feature fusion module to capture histopathological information across multiple scales, enhancing the generation of fine details.
\item The proposed method achieves state-of-the-art performance on the TCGA-NSCLC datasets.
\end{itemize}

\section{Methods}

The overall structure of Diffusion-FFPE is illustrated in Fig. \ref{overview}. The network primarily consists of the generator $G$ and the discriminator $D$. To leverage prior knowledge from pre-trained models, we adopt a pre-trained one-step diffusion model with trainable LoRA adapters as the generator and employ a pre-trained vision model as the backbone of discriminator. During training, the generator is fine-tuned with LoRA adapters using adversarial optimization objectives, while a multi-scale feature fusion (MFF) module is introduced to enhance the generation of fine details. In the inference phase, the fine-tuned generator translates FF images into the FFPE image domain.

\subsection{The Generator with a Pre-trained Diffusion Model}
\label{generator}
Inspired by img2img-turbo\cite{parmar2024one}, we adopt sd-turbo as the generator $G$, which can synthesize realistic images from a text prompt in a single network evaluation. It consists of a VAE encoder $\mathcal{E}$, a VAE decoder $\mathcal{D}$, a UNet $\epsilon_\theta$, and a text encoder $\tau_\theta$. First, the VAE encoder $\mathcal{E}$ extracts features from the image $x$ and converts them into the latent representation $z_x = \mathcal{E}(x)$. The text encoder $\tau_\theta$ converts the text prompt $c_Y$ into the text representation $\tau_\theta(c_Y)$. Both the image latent representation $z_x$ and the text representation $\tau_\theta(c_Y)$ are then fed into the UNet to predict the noise $\epsilon=\epsilon_\theta(z_x, \tau_\theta(c_Y))$. Next, the denoised latent representation $z_y=s.step(z_x, \epsilon)$ is computed using a schedule $s$. Finally, the VAE decoder $\mathcal{D}$ decodes the latent representation $z_y$ to obtain the translated image $y=\mathcal{D}(z_y)$.

To enable the pre-trained model to learn the distribution of histopathological images, we add trainable LoRA adapters to each layer of the VAE and UNet. Additionally, skip connections between the VAE encoder and decoder are implemented to preserve image details and mitigate information loss during encoding. Zero convolution layers with weights initialized to zeros are employed to facilitate learning in a residual manner.

\subsection{The Multi-scale Feature Fusion Module}
The MFF module is designed to enable the model to focus on smaller regions within an image. First, we divide the image $x$ into multiple small patches $\{x_i\}_{i=1}^N$. A global VAE encoder extracts global feature $F$ from the image $x$ and a local VAE encoder extracts local features $\{f_i\}_{i=1}^N$ from small patches. Notably, $F$ and $\{f_i\}_{i=1}^N$ denote the intermediate feature before being transformed into the latent space.

The MFF module integrates the intermediate features \( F^l\) and \( \{f_i^l\}_{i=1}^N \) from each layer $l$ of the global and local VAE encoders. Specifically, the local features are squared based on their positions in the original image to match the dimensions of the global feature. The MFF module fuses these features by summing the local and global features to obtain a fused representation \( F_{fused}^l \) for each layer $l$:

\begin{equation}
    F_{fused}^l=F^l + squared(\{f_i^l\}_{i=1}^N).
    \label{fused}
\end{equation}

The fused features from each layer are subsequently forwarded to the VAE decoder $\mathcal{D}$ via skip connections: 
\begin{equation}
    y^l=\mathcal{D}^l(y^{l-1}) + z_\theta(F_{fused}^l),
\end{equation}
where \( y^l \) denotes the feature map from layer \( l \) in the VAE decoder and \(z_\theta\) denotes the zere convolution layer.

The last fused feature \( F_{fused}^L \) is transformed into the latent variable \( z_x \) by the last layer of the VAE encoder $\mathcal{E}_{last}$ and then forwarded to the U-Net:

\begin{equation}
    z_x=\mathcal{E}_{last}(F_{fused}^L).
\end{equation}

\subsection{The Discriminator with a Pre-trained Vision Model}
\label{discriminator}
To increase the training efficiency, we use a pre-trained visual model $\Theta$ for histopathology as the backbone of the discriminator $D$. We adopt the vision-aided GAN approach\cite{kumari2022ensembling}, where the weights of the pre-trained visual model are kept fixed, followed by a small classifier head $C$:
\begin{equation}
    D(x)=C(\Theta(x)).
\end{equation}

\subsection{Adversarial Learning Objective}
Diffusion-FFPE is trained based on the formulation of CycleGAN\cite{zhu2017unpaired}, which consists of two mapping functions: $G_Y(x, c_Y): X \rightarrow Y$ and $G_X(y, c_X): Y \rightarrow X$. The $G_X$ and $G_Y$ networks have identical structures and share UNet weights, but they utilize different VAE encoders and decoders. Additionally, they receive different texts $c_X$ and $c_Y$ to perform their respective translation tasks.

To apply adversarial losses to the mapping functions $G_X$ and $G_Y$, we employ discriminators $D_X$ and $D_Y$ respectively. This ensures that the generated output images match the target domain. The adversarial objective $L_{adv}$ is defined as:

\begin{equation}
    \begin{aligned}
    L_{adv} &= E_y[logD_Y(y)] \\
    &+ E_x[log(1-D_Y(G_Y(x,c_Y)))] \\
    &+ E_x[logD_X(x)] \\
    &+ E_y[log(1-D_X(G_X(y,c_X)))].
    \end{aligned}
\end{equation}

The cycle consistency loss is necessary for maintaining content consistency between FF images and FFPE images. It ensures that when an FF image $x$ is mapped through $G_Y$ to generate an FFPE image $G_Y(x, c_Y)$ and then mapped back through $G_X$, it should return to the original image $x \approx G_X(G_Y(x, c_Y), c_X)$. Besides, the reconstruction loss $L_{rec}$ is used to measure the similarity between the images:

\begin{equation}
    \begin{aligned}
    L_{cyc} &= E_x[L_{rec}(G_X(G_Y(x, c_Y), c_X), x)] \\
            &+ E_y[L_{rec}(G_Y(G_X(y, c_X), c_Y), y)],
    \end{aligned}
\end{equation}

\begin{equation}
L_{rec}=\lambda_1 L_1+\lambda_p L_p.
\end{equation}

The reconstruction loss is defined as a linear combination of the $L_1$ norm and the Learned Perceptual Image Patch Similarity (LPIPS) $L_p$ weighted by  parameters $\lambda_1$ and $\lambda_p$. Additionally, an identity regularization loss $L_{idt}$ is employed to ensure that the generator does not alter images from the target domain:

\begin{equation}
    \begin{aligned}
    L_{idt} &= E_x[L_{rec}(G_X(x, c_X), x)] \\
            &+ E_y[L_{rec}(G_Y(y, c_Y), y)].
    \end{aligned}
\end{equation}

In general, the overall optimization objective $L_{total}$ is represented as follows, weighted by the hyperparameters $\lambda_{adv}$, $\lambda_{cyc}$ and $\lambda_{idt}$:
\begin{equation}
    L_{total} = \lambda_{adv}L_{adv} + \lambda_{cyc}L_{cyc} + \lambda_{idt}L_{idt}.
\end{equation}

\section{Experiments}
\subsection{Datasets and Implementation Details}
We conduct experiments on the TCGA non-small cell lung cancer (TCGA-NSCLC) dataset. The WSIs are cropped into multiple 512x512 patches at 20x magnification. We use a subset consisting of 50,000 pairs of FF and FFPE patches for training, along with 2,000 FFPE images for validation and 10,000 FFPE images for final evaluation. The Frechet Inception Distance (FID) and Kernel Inception Distance (KID) metrics are used to measure whether the generated images match the FFPE data distribution.

 Diffusion-FFPE is implemented in PyTorch and trained for 50,000 steps with a batch size of 1. We employ CONCH \cite{lu2024visual} as the discriminator, owing to its demonstrated performance across a range of downstream tasks. We use the Adam optimizer with an initial learning rate of \(5 \times 10^{-6}\), \(\beta_1 = 0.9\), and \(\beta_2 = 0.999\). We set \(\lambda_1 = 1\) for both \(L_{idt}\) and \(L_{cyc}\), with \(\lambda_p = 10\) in \(L_{cyc}\) and \(\lambda_p = 1\) in \(L_{idt}\). The weights for the total loss \(L_{\text{total}}\) are \(\lambda_{\text{adv}} = 0.5\), \(\lambda_{\text{cyc}} = 1\), and \(\lambda_{\text{idt}} = 1\). For the text prompts, \(c_X\) is "frozen section," and \(c_Y\) is "paraffin section."

\subsection{Comparison Experiments}
\begin{figure}[htbp]
    \centering
    \includegraphics[width=0.46\textwidth]{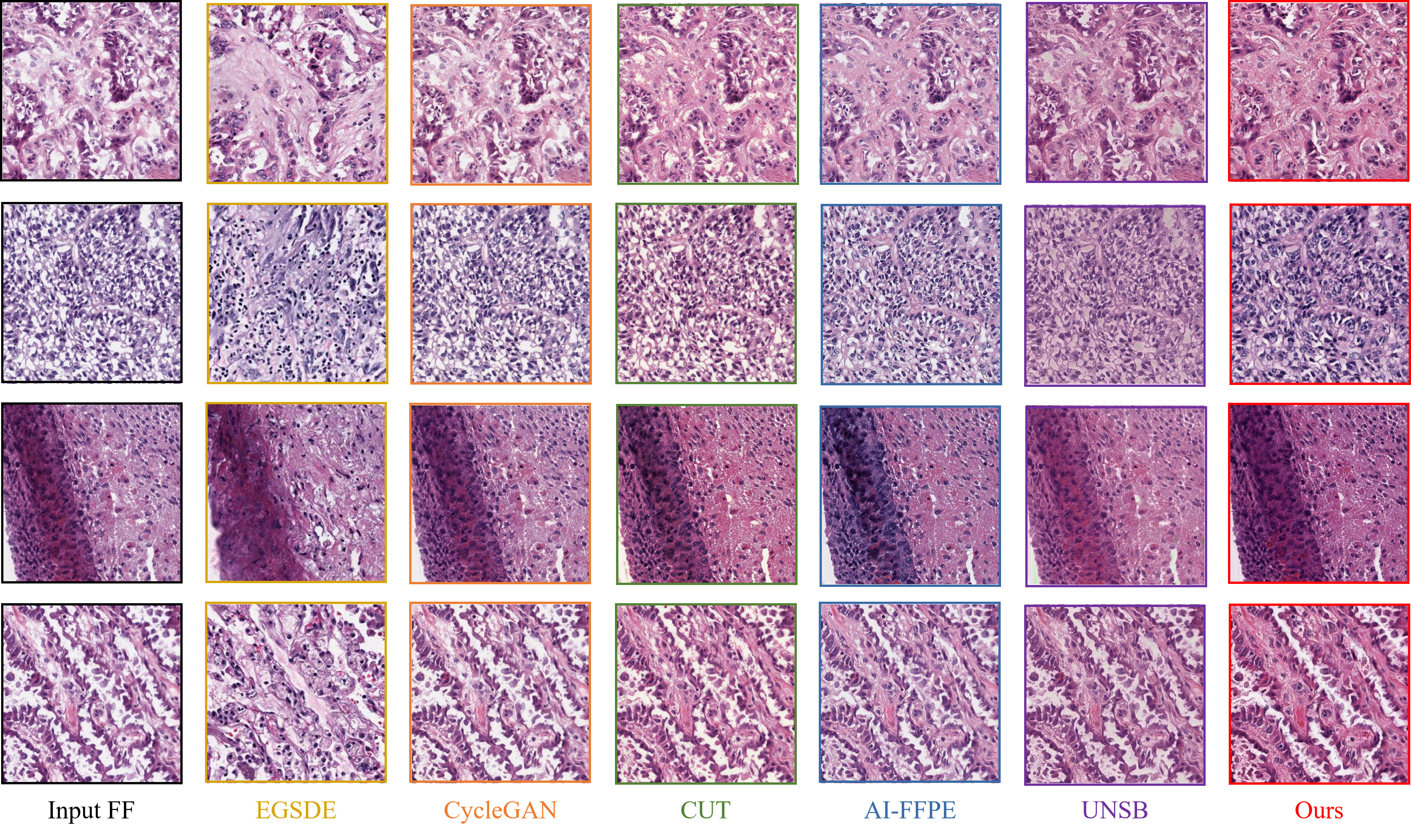}
    \caption{Visualization results of comparison experiments.}
    \label{visual}
\end{figure}

We compare our method with other GAN-based and diffusion-based approaches. As shown in Table \ref{main_results}, our method achieves an FID score of 15.78 and a KID score of 8.17$\times$10$^{-3}$, outperforming the competing methods. Fig. \ref{visual} illustrates that our translated images exhibit a more distinct FFPE style compared to other methods, enabling clearer differentiation between tissue and blank areas and effectively reducing artifacts within tissue regions.

\begin{table}[htbp]
\renewcommand{\arraystretch}{1}
\centering
\caption{Comparison Experiments on TCGA-NSCLC Datasets}
\footnotesize
\begin{tabular}{c c c}
\hline
\textbf{Model} & \textbf{FID} & \textbf{KID ($\times10^3$)} \\
\hline
 CycleGAN\cite{zhu2017unpaired} & 34.85 & 26.00 \\
 CUT\cite{park2020contrastive}  & 33.86 & 22.39 \\
 AI-FFPE\cite{ozyoruk2022deep}   & 25.90  & 17.42 \\
 EGSDE\cite{zhao2022egsde} & 62.69 & 61.28 \\
 UNSB\cite{kim2023unpaired} & 36.37 & 26.94\\
 Diffusion-FFPE (Ours) & \textbf{15.78} & \textbf{8.17}\\
\hline
\label{main_results}
\end{tabular}
\end{table}

\begin{table}[htbp]
\renewcommand{\arraystretch}{1}
\centering
\caption{The Ablation Analysis of Diffusion-FFPE}
\footnotesize
\begin{tabular}{c c c c c}
\hline
\textbf{Generator}& \textbf{Discirminator} & \textbf{MFF Module} & \textbf{FID} & \textbf{KID} ($\times10^3$)\\
\hline
\emph{initalized} & CONCH & \emph{each layer} & 42.36 & 30.46 \\
\emph{pre-trained} & CONCH & \emph{each layer} & \textbf{15.78} & \textbf{8.17}\\
\hline
\emph{pre-trained} & PatchGAN & \emph{each layer} & 26.85 & 17.54\\
\emph{pre-trained} & CLIP & \emph{each layer} & 19.47 & 10.37\\
\emph{pre-trained} & CONCH & \emph{each layer} & \textbf{15.78} & \textbf{8.17}\\
\hline
\emph{pre-trained} & CONCH & \emph{not used} & 18.15 & 9.75\\
\emph{pre-trained} & CONCH & \emph{last layer} & 16.24 & 8.97\\
\emph{pre-trained} & CONCH & \emph{each layer} & \textbf{15.78} & \textbf{8.17}\\
\hline
\label{ablation}
\end{tabular}
\end{table}

\begin{figure}[htbp]
    \centering
    \includegraphics[width=0.45\textwidth]{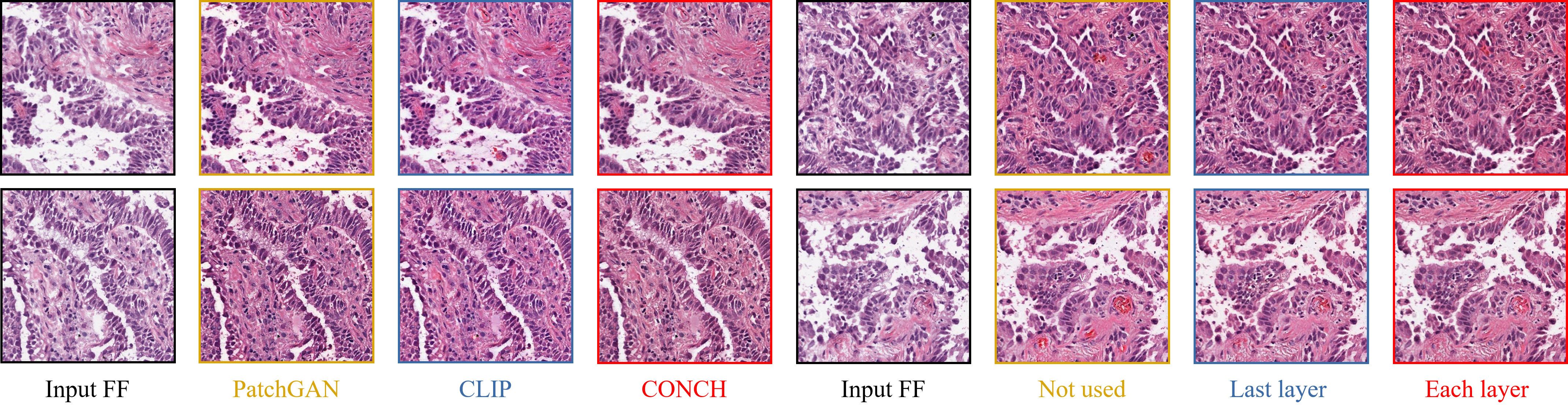}
    \caption{Visualization results of ablation study.}
    \label{fig_ablation}
\end{figure}

\subsection{Ablation Study}
\subsubsection{The Impact of Pre-trained Weights in Generator}
To assess the impact of prior knowledge on the generation of FFPE images, we train the model with randomly initialized weights in the VAE and UNet components of the generator. As shown in Table \ref{ablation}, the model with pre-trained weights significantly outperforms the model with randomly initialized weights, demonstrating that pre-trained weights are essential for high-quality generated results.


\subsubsection{The Impact of Discirminator}
We further evaluate the discriminator by conducting experiments with randomly initialized versions of PatchGAN \cite{isola2017image}, CLIP \cite{radford2021learning}, and CONCH \cite{lu2024visual}. As illustrated in Table \ref{ablation}, the discriminator leveraging CONCH as the backbone achieves superior performance compared to other configurations. Fig. \ref{fig_ablation} shows that images generated with CLIP as the discriminator tend to display unintended red-highlighted regions, a limitation that CONCH effectively mitigates.



\subsubsection{The Impact of Multi-scale Feature Fusion Module}

We conducted three experiments to evaluate the MFF module: one without MFF, one with fusion on the last layer, and one with fusion across all VAE encoder layers. As presented in Table \ref{ablation}, incorporating MFF modestly improves model performance. Fig. \ref{fig_ablation} shows that red artifacts appear in the images without MFF, while applying MFF to the last or all encoder layers effectively reduces these artifacts. This improvement results from the local VAE encoder's focus on localized features, minimizing red cell artifacts in the generated images.


\section{Conclusion}
This paper proposes Diffusion-FFPE, a method for FF-to-FFPE histopathological image translation using pre-trained models. Specifically, a pre-trained one-step diffusion model serves as the generator, leveraging its generative prior effectively, while a pre-trained histopathology vision model acts as the discriminator backbone to enhance GAN training. We further introduce a multi-scale feature fusion module to refine detail translation by focusing on smaller image regions. This approach is also adaptable to other medical image translation tasks, such as CT-to-PET.

\section*{acknowledgement}
The work is supported by National Natural Science Foundation of China (No. 82430062), Shenzhen Engineering Research Centre (No. XMHT20230115004), Science and Technology Research Program of Shenzhen City (No. KCXFZ20201221173207022), and Jilin Fuyuan Guan Food Group Co., Ltd. In.


\bibliographystyle{IEEEtran}
\begin{spacing}{0.8}
    \bibliography{myreference.bib}

\begin{thebibliography}{10}
\providecommand{\url}[1]{#1}
\csname url@samestyle\endcsname
\providecommand{\newblock}{\relax}
\providecommand{\bibinfo}[2]{#2}
\providecommand{\BIBentrySTDinterwordspacing}{\spaceskip=0pt\relax}
\providecommand{\BIBentryALTinterwordstretchfactor}{4}
\providecommand{\BIBentryALTinterwordspacing}{\spaceskip=\fontdimen2\font plus
\BIBentryALTinterwordstretchfactor\fontdimen3\font minus \fontdimen4\font\relax}
\providecommand{\BIBforeignlanguage}[2]{{%
\expandafter\ifx\csname l@#1\endcsname\relax
\typeout{** WARNING: IEEEtran.bst: No hyphenation pattern has been}%
\typeout{** loaded for the language `#1'. Using the pattern for}%
\typeout{** the default language instead.}%
\else
\language=\csname l@#1\endcsname
\fi
#2}}
\providecommand{\BIBdecl}{\relax}
\BIBdecl

\bibitem{rogers1987accuracy}
C.~Rogers, E.~Klatt, and P.~Chandrasoma, ``Accuracy of frozen-section diagnosis in a teaching hospital.'' \emph{Archives of pathology \& laboratory medicine}, vol. 111, no.~6, pp. 514--517, 1987.

\bibitem{jaafar2006intra}
H.~Jaafar, ``Intra-operative frozen section consultation: concepts, applications and limitations,'' \emph{The Malaysian journal of medical sciences: MJMS}, vol.~13, no.~1, p.~4, 2006.

\bibitem{trejo2015accuracy}
H.~E. Trejo~Bittar, P.~Incharoen, A.~D. Althouse, and S.~Dacic, ``Accuracy of the iaslc/ats/ers histological subtyping of stage i lung adenocarcinoma on intraoperative frozen sections,'' \emph{Modern Pathology}, vol.~28, no.~8, pp. 1058--1063, 2015.

\bibitem{falahkheirkhah2022generative}
K.~Falahkheirkhah, T.~Guo, M.~Hwang, P.~Tamboli, C.~G. Wood, J.~A. Karam, K.~Sircar, and R.~Bhargava, ``A generative adversarial approach to facilitate archival-quality histopathologic diagnoses from frozen tissue sections,'' \emph{Laboratory Investigation}, vol. 102, no.~5, pp. 554--559, 2022.

\bibitem{ozyoruk2022deep}
K.~B. Ozyoruk, S.~Can, B.~Darbaz, K.~Ba{\c{s}}ak, D.~Demir, G.~I. Gokceler, G.~Serin, U.~P. Hacisalihoglu, E.~Kurtulu{\c{s}}, M.~Y. Lu \emph{et~al.}, ``A deep-learning model for transforming the style of tissue images from cryosectioned to formalin-fixed and paraffin-embedded,'' \emph{Nature Biomedical Engineering}, vol.~6, no.~12, pp. 1407--1419, 2022.

\bibitem{fan2022fast}
L.~Fan, A.~Sowmya, E.~Meijering, and Y.~Song, ``Fast ff-to-ffpe whole slide image translation via laplacian pyramid and contrastive learning,'' in \emph{International Conference on Medical Image Computing and Computer-Assisted Intervention}.\hskip 1em plus 0.5em minus 0.4em\relax Springer, 2022, pp. 409--419.

\bibitem{li2023st}
Z.~Li, Y.~Lin, Y.~Wang, Z.~Fang, H.~Bian, R.~Hu, X.~Li, and Y.~Zhang, ``St-mksc: The ff-ffpe stain transfer based on multiple key structure constraint,'' in \emph{2023 IEEE 20th International Symposium on Biomedical Imaging (ISBI)}.\hskip 1em plus 0.5em minus 0.4em\relax IEEE, 2023, pp. 1--5.

\bibitem{rombach2022high}
R.~Rombach, A.~Blattmann, D.~Lorenz, P.~Esser, and B.~Ommer, ``High-resolution image synthesis with latent diffusion models,'' in \emph{Proceedings of the IEEE/CVF conference on computer vision and pattern recognition}, 2022, pp. 10\,684--10\,695.

\bibitem{zhang2023customized}
K.~Zhang and D.~Liu, ``Customized segment anything model for medical image segmentation,'' \emph{arXiv preprint arXiv:2304.13785}, 2023.

\bibitem{lu2024visual}
M.~Y. Lu, B.~Chen, D.~F. Williamson, R.~J. Chen, I.~Liang, T.~Ding, G.~Jaume, I.~Odintsov, L.~P. Le, G.~Gerber \emph{et~al.}, ``A visual-language foundation model for computational pathology,'' \emph{Nature Medicine}, vol.~30, no.~3, pp. 863--874, 2024.

\bibitem{kumari2022ensembling}
N.~Kumari, R.~Zhang, E.~Shechtman, and J.-Y. Zhu, ``Ensembling off-the-shelf models for gan training,'' in \emph{Proceedings of the IEEE/CVF conference on computer vision and pattern recognition}, 2022, pp. 10\,651--10\,662.

\bibitem{parmar2024one}
G.~Parmar, T.~Park, S.~Narasimhan, and J.-Y. Zhu, ``One-step image translation with text-to-image models,'' \emph{arXiv preprint arXiv:2403.12036}, 2024.

\bibitem{hu2021lora}
E.~J. Hu, Y.~Shen, P.~Wallis, Z.~Allen-Zhu, Y.~Li, S.~Wang, L.~Wang, and W.~Chen, ``Lora: Low-rank adaptation of large language models,'' \emph{arXiv preprint arXiv:2106.09685}, 2021.

\bibitem{zhu2017unpaired}
J.-Y. Zhu, T.~Park, P.~Isola, and A.~A. Efros, ``Unpaired image-to-image translation using cycle-consistent adversarial networks,'' in \emph{Proceedings of the IEEE international conference on computer vision}, 2017, pp. 2223--2232.

\bibitem{park2020contrastive}
T.~Park, A.~A. Efros, R.~Zhang, and J.-Y. Zhu, ``Contrastive learning for unpaired image-to-image translation,'' in \emph{Computer Vision--ECCV 2020: 16th European Conference, Glasgow, UK, August 23--28, 2020, Proceedings, Part IX 16}.\hskip 1em plus 0.5em minus 0.4em\relax Springer, 2020, pp. 319--345.

\bibitem{zhao2022egsde}
M.~Zhao, F.~Bao, C.~Li, and J.~Zhu, ``Egsde: Unpaired image-to-image translation via energy-guided stochastic differential equations,'' \emph{Advances in Neural Information Processing Systems}, vol.~35, pp. 3609--3623, 2022.

\bibitem{kim2023unpaired}
B.~Kim, G.~Kwon, K.~Kim, and J.~C. Ye, ``Unpaired image-to-image translation via neural schr\"{o}dinger bridge,'' \emph{arXiv preprint arXiv:2305.15086}, 2023.

\bibitem{isola2017image}
P.~Isola, J.-Y. Zhu, T.~Zhou, and A.~A. Efros, ``Image-to-image translation with conditional adversarial networks,'' in \emph{Proceedings of the IEEE conference on computer vision and pattern recognition}, 2017, pp. 1125--1134.

\bibitem{radford2021learning}
A.~Radford, J.~W. Kim, C.~Hallacy, A.~Ramesh, G.~Goh, S.~Agarwal, G.~Sastry, A.~Askell, P.~Mishkin, J.~Clark \emph{et~al.}, ``Learning transferable visual models from natural language supervision,'' in \emph{International conference on machine learning}.\hskip 1em plus 0.5em minus 0.4em\relax PMLR, 2021, pp. 8748--8763.

\end{thebibliography}
\end{spacing}


\end{document}